\documentclass[letterpaper]{article} 
\usepackage{aaai2026}  
\usepackage{times}  
\usepackage{helvet}  
\usepackage{courier}  
\usepackage[hyphens]{url}  
\usepackage{graphicx} 
\usepackage{natbib}  
\usepackage{caption} 
\usepackage{multibib}

\usepackage[T1]{fontenc}
\input{glyphtounicode}
\newcites{app}{Appendix References}
\newcommand{\rot}[1]{\rotatebox{90}{\parbox{4.2cm}{\raggedright #1}}}
\makeatletter
\newcommand{\blindfootnote}[1]{%
  \bgroup
  \renewcommand{\@makefntext}[1]{\noindent ##1}%
  \let\thefootnote\relax
  \footnotetext{#1}%
  \egroup
}
\makeatother
\nocopyright

\title{No One to Blame: A Framework of Constitutive AI Unaccountability}

\author{
    Long Hoang Nguyen\textsuperscript{\rm 1},
    Eva Sp{\"a}the\textsuperscript{\rm 2},
    Sebastian Lins\textsuperscript{\rm 2},
    Ali Sunyaev\textsuperscript{\rm 1}
}
\affiliations{
    \textsuperscript{\rm 1}Technical University of Munich\\
    \textsuperscript{\rm 2}University of Kassel\\
    long.hoang.nguyen@tum.de, eva.spaethe@uni-kassel.de, sebastian.lins@uni-kassel.de, sunyaev@tum.de
}

\begin{document}

\maketitle

\begin{abstract}
The increasing deployment of autonomous, agentic AI systems challenges traditional accountability mechanisms. Existing research predominantly frames AI accountability gaps as barriers that can be overcome through better standards, transparency, and institutional reform. We argue that this framing is insufficient: certain configurations of actors, systems, and institutions render AI accountability conceptually unachievable regardless of effort. We introduce the concept of constitutive AI unaccountability to capture these configurations. Through a three-stage qualitative study comprising a concept-centric literature analysis, a secondary analysis of 27 expert interviews with AI professionals from technical, legal, and sociotechnical backgrounds, and an illustrative framework application to the open-source agentic AI system OpenClaw, we identify nine categories and 20 themes of constitutive AI unaccountability. These are organized across structural, technological, and normative clusters and reinforce one another through eight directed interdependencies. Our framework is operationalized as a diagnostic instrument of 20 questions, which detected 17 of 20 conditions when applied to OpenClaw, including an inverted anthropomorphism configuration in which the AI agent was the only identifiable actor. We contribute a reframing of AI unaccountability as a constitutive property of sociotechnical systems, an extension of the four barriers to accountability, and a practical instrument for identifying accountability voids in specific AI deployments.
\end{abstract}

\begin{links}
     \link{Supplementary Materials}{https://osf.io/jtqkr}
\end{links}

\section{Introduction}\blindfootnote{Accepted at the Ninth AAAI/ACM Conference on AI, Ethics, and Society (AIES 2026), Malm\"{o}, Sweden, October 12--14, 2026.}

As contemporary AI systems are increasingly deployed in everyday life, including highly sensitive domains such as recruitment and credit scoring, the ability to hold actors accountable for their outcomes becomes both more important and harder to achieve. Accountability provides the institutional mechanism through which harms can be addressed, standards enforced, and trust maintained \cite{Bovens_2007}. In the context of AI, this translates to the obligation of actors involved in the design, development, and use of AI systems to explain and justify the system's decisions, outcomes, and impacts to relevant fora (e.g., regulatory bodies), and to face consequences when those explanations prove inadequate \cite{Wieringa_2020}.

However, as AI systems grow in complexity, autonomy, and societal reach, the conditions under which accountability as a mechanism can function become increasingly difficult to satisfy. On the one hand, there is a growing number of actors that are now being held accountable, from individual developers and the organizations that deploy the systems to the AI systems themselves \cite{Banteka_2020, Wieringa_2020}. On the other hand, agentic AI systems further complicate accountability because their decisions emerge from autonomous processes that no single human actor oversees or controls \cite{Banteka_2020, Turner_2018}. Contemporary agentic AI systems such as OpenClaw \cite{Steinberger_2025} autonomously execute complex, multi-step workflows over extended time horizons with minimal human intervention \cite{Acharya_etal_2025, Wang_etal_2024}. Not only do those agentic AI deployments strain traditional accountability mechanisms, they may also render them fundamentally inapplicable. Together, these dynamics risk producing a situation in which accountability is either distributed across so many actors that none is answerable, or directed at actors that cannot bear it in any meaningful sense. In the end, no one is effectively held to account for harm. 

Prior research has predominantly focused on identifying who is accountable by assigning liability based on an actor's degree of control or involvement in AI-driven outcomes \cite{Cooper_etal_2022, Nissenbaum_1996, Shneiderman_2020}. Extant literature identifies four recurring barriers that impede this attribution: the diffusion of responsibility, technical errors, scapegoating, and the severing of ownership from liability \cite{Cooper_etal_2022, Nissenbaum_1996, Xia_etal_2024}. Such work frequently treats accountability failures as obstacles that can be overcome through better standards, auditing, and institutional reform. We argue that this barrier-centric framing is insufficient because some conditions do not stem from removable obstacles but from AI systems lacking capacities that accountability presupposes \cite{Gunkel_2020, Ma_Su_2025}. For instance, the attribution of legal accountability to AI agents remains largely unenforceable given that AI systems cannot be sued, imprisoned, or made to experience remorse \cite{Gunkel_2020, Ma_Su_2025}. Some actors even actively benefit from maintaining these conditions rather than resolving them \cite{Vesa_Tienari_2022}. Identifying where accountability is unachievable is necessary to prevent its misattribution to actors that cannot meet it \cite{Elish_2019, Martin_2019} and to direct governance efforts where they can be effective. Therefore, we propose shifting the focus from how accountability barriers can be overcome to the conditions under which meaningful accountability cannot be achieved (i.e., \textbf{UN}accountability). We ask: \textit{Under what conditions can AI accountability not be achieved within a sociotechnical configuration, regardless of effort?}

We adopt a three-stage qualitative approach. First, we conduct a concept-centric literature analysis \cite{Pare_etal_2015, Webster_Watson_2002} to develop an initial framework of constitutive AI unaccountability conditions. Second, we refine and extend the framework through a secondary analysis of 27 expert interviews originally collected for a broader AI accountability research project. Third, we apply the resulting framework to OpenClaw \cite{Steinberger_2025} to illustrate its diagnostic capacity.

Our study makes three contributions. First, we develop a framework of nine categories and 20 themes of constitutive AI unaccountability, organized across structural, technological, and normative clusters. Our framework extends the four barriers to accountability by identifying conditions absent from prior work, mapping their interdependencies, and revealing asymmetries between academic and practitioner understanding. Second, we operationalize the framework as a 20-question diagnostic instrument for direct use by practitioners, regulators, and auditors assessing constitutive unaccountability in specific AI deployments. Third, we apply the instrument to OpenClaw, surfacing 17 of 20 conditions including an inverted anthropomorphism configuration in which the AI agent operated under a constructed human identity while its operator remained unidentifiable.

The paper is structured as follows. Section 2 introduces AI accountability and the theoretical foundation of AI unaccountability. Section 3 outlines our research approach. Section 4 presents the framework of constitutive AI unaccountability conditions, the interdependencies between them, and the framework application to OpenClaw. Section 5 discusses implications for research and practice, followed by limitations and future research. Section 6 concludes the paper.

\section{Background}
\subsection{AI Accountability}
Traditional AI systems, including supervised classifiers or early generative models, function as passive tools that execute strictly defined tasks within controlled boundaries \cite{Acharya_etal_2025}. In contrast, agentic AI systems refer to a class of highly autonomous, adaptable systems designed to pursue complex, open-ended goals over extended time horizons with minimal human intervention. Contemporary agentic AI systems frequently rely on large language models (LLMs) as reasoning engines to turn language understanding into autonomous action \cite{Wang_etal_2024}. This enables capabilities such as autonomous code generation and multi-step task execution seen in agents like OpenClaw \cite{Steinberger_2025}, but it also means that the AI system is evolving from a predictable tool into an autonomous actor whose behavior may no longer be fully controllable. This shift has renewed attention to the conditions under which actors behind systems can be held accountable.

Accountability generally refers to ``a relationship between an actor and a forum, in which the actor has the obligation to explain and to justify his or her conduct, the forum can pose questions and pass judgment, and the actor may face consequences'' \cite[p. 450]{Bovens_2007}. \citet{Wieringa_2020} contextualizes this definition by understanding AI accountability as creating an account of a sociotechnical AI system with multiple actors (e.g., developers and users) who must explain and justify the system's design, use, decisions, and outcomes to various kinds of fora (e.g., regulatory bodies). Actors can be held accountable for a specific aspect of the system or for the entire system.

\subsection{From Barriers to Constitutive AI Unaccountability}
Extant AI accountability research has identified four recurring barriers to holding actors accountable for AI-driven outcomes: (1) the problem of many hands, where responsibility is diffused across multiple contributors; (2) bugs, where adverse outcomes are attributed to unintentional technical errors; (3) scapegoating, where blame is redirected to convenient targets; and (4) ownership without liability, where actors claim the benefits of AI while disclaiming responsibility for its harms \cite{Cooper_etal_2022, Nissenbaum_1996}. Previous studies have largely treated these barriers as obstacles that can be overcome by, for instance, improving standards, auditing practices, and engaging in institutional reform \cite[e.g.,][]{Ananny_Crawford_2018, Cooper_etal_2022}. We argue that this barrier-centric framing is insufficient to examine who can be held accountable for three reasons.

First, these barriers have consistently been studied in isolation. \citet{Nissenbaum_1996} introduces each barrier individually. \citet{Cooper_etal_2022} revisit them for algorithmic systems following the same structure, only noting informally that bugs and scapegoating can `see-saw' (p.~871). Similarly, \citet{Xia_etal_2024} elaborate on barriers in the generative AI context without examining how they interact. Studying barriers in isolation risks missing how they compound into an accountability failure that no single barrier can explain.

Second, not all conditions that undermine AI accountability are obstacles that can be overcome or removed. AI systems cannot be treated as legal persons that can be sued, biological beings that can be punished, or spiritual persons that can experience remorse \cite{Gunkel_2020, Ma_Su_2025}. Despite proposals to map legal liability directly onto AI systems \cite[e.g.,][]{Banteka_2020, Turner_2018}, these conditions are not obstacles that better engineering or regulation can remove. AI systems lack the very capacities that accountability presupposes of its bearers, risking the manifestation of situations where no actor is held accountable in practice.

Third, the focus on barriers treats accountability failures as deficiencies that need to be corrected. Yet recent work demonstrates that some actors actively benefit from and maintain these conditions of unaccountability rather than seeking to resolve them. For example, \citet{Vesa_Tienari_2022} argue that organizations rationalize and exploit the absence of accountability as a strategic resource.

These three observations emphasize limitations of the existing research and point to the problem that the prevalent focus on barriers cannot adequately capture the accountability challenges specific to AI. Ultimately, AI accountability may fail not because obstacles stand in the way but because the conditions under which AI systems are built, deployed, and governed make accountability unachievable. We propose the term \textit{constitutive AI unaccountability} for this reframing. 

To overcome identified limitations, we examine under what conditions accountability cannot be achieved within a sociotechnical configuration, regardless of effort. This reframing builds on the relational understanding of accountability, which presupposes an actor capable of occupying the answering role in a dispute with a forum. Following \citet{Searle_1995}, occupying such a role is an institutional status, comparable to being a company director. Expertise alone makes no one a director. The status exists only through the constitutive rules of an institution, which create the very possibility of holding it and confer it on those who are expected to be capable of performing the role. The same holds for being accountable. Institutions confer this status on actors expected to justify conduct, submit to judgment, and face consequences. Constitutive AI unaccountability thus refers to sociotechnical configurations in which no actor satisfying these presuppositions is available, leaving the accountability relationship with nothing to attach to. This relationship is constitutive rather than causal. The configurations do not merely cause accountability to fail but constitute the conditions under which it cannot be achieved \cite{Haslanger_2003}. AI unaccountability is thus constituted by the configuration of actors, systems, and institutions, not by the technology alone \cite{Nabben_2024, Suchman_2007}. Because unaccountability is constituted at the level of the configuration, no effort within a persisting configuration can achieve accountability. What varies is whether and how deeply a configuration must change for accountability to become possible.

We further draw on two concepts to operationalize AI unaccountability. First, accountability sinks \cite{Davies_2024} describe systems, such as a dense bureaucracy or a black-box algorithm, that is structurally designed to absorb blame so that no actor can be identified as answerable. Second, rationalized unaccountability \cite{Vesa_Tienari_2022} captures the organizational ideology through which actors exploit such structures, framing AI as objective and independent of human influence to shield themselves from consequences.

\section{Methodology}
We applied a three-stage qualitative research approach. First, we conducted a concept-centric literature analysis \cite{Pare_etal_2015, Webster_Watson_2002} to develop an initial framework of constitutive AI unaccountability. Second, we conducted a secondary analysis of 27 expert interviews, originally collected for a broader AI accountability research project, to refine and extend the framework with practitioner perspectives. Third, we applied the resulting framework to the case of OpenClaw to illustrate its diagnostic capacity.

\subsection{Stage 1: Concept-Centric Literature Analysis}
We conducted a theoretical review \cite[cf.][]{Pare_etal_2015}, organized concept-centrically \cite[cf.][]{Webster_Watson_2002}, treating each paper as evidence for one or more conditions. 

We searched the Scopus database on March 10, 2026 using the query \textit{(``Artificial Intelligence'' OR ``AI'' OR ``algorithm*'') AND (``unaccountab*'')}, limiting results to peer-reviewed English-language publications. This search returned 472 results. After screening titles and abstracts for relevance to constitutive AI unaccountability as a substantive topic rather than a passing mention, 14 papers remained. We then conducted forward and backward searches \cite{Webster_Watson_2002}, yielding 31 candidate papers in total. Full-text screening led to the exclusion of 16 papers that did not substantively address constitutive conditions of AI unaccountability, resulting in 15 papers selected for analysis.

We applied a template analysis approach \cite[cf.][]{King_2012}. After inductively coding the first two papers, we arrived at an initial coding template of seven conditions: opacity and unexplainability, incapacity to bear consequences, responsibility diffusion, temporal asymmetry, anthropomorphism and scapegoating, ideological rationalization, and emergent systemic behavior. We then iteratively applied and refined this template across the remaining 13 papers, deductively coding each relevant text passage by assigning it to a condition and noting underlying mechanisms. Where text passages did not fit the existing template, we flagged them for potential new conditions. Through iterative refinement, including merging overlapping conditions to themes and later categories, and splitting them where distinct mechanisms emerged, we arrived at nine categories comprising 19 themes of conditions for unaccountability (Table~\ref{tab:framework}), which were assigned to 315 text segments in total.

\subsection{Stage 2: Expert Interviews}
To refine and extend identified conditions, we conducted a secondary analysis of 27 expert interviews, which we originally collected for a broader research project on AI accountability. Secondary analysis of qualitative data is an established methodological practice when the original data collection aligns closely with the new research question \cite[cf.][]{Heaton_2008}. In our case, the broader research project investigates how AI accountability is conceptualized across disciplinary perspectives, making the interview transcripts directly relevant to identifying the conditions under which accountability cannot be achieved.

The 27 interviews were conducted with AI professionals purposefully sampled \cite[cf.][]{Patton_2014} to represent three perspectives on AI accountability: technical (13; e.g., AI engineer), legal (7; e.g., Professor for Law), and sociotechnical (7; e.g., AI researcher - responsible AI). We sampled participants based on two criteria: (1) currently working in a relevant AI-related position (e.g., data scientist), and (2) multiple years of professional AI-related experience. Participants were recruited through LinkedIn and personal contacts, held advanced degrees (10 PhDs, 16 Master's, 1 Bachelor's), and had 2.5 to 20 years of professional AI experience (average: 6 years). The interviews lasted 48 minutes on average (for participant details see the supplementary materials).

During the interview analysis, several refinements to the template emerged. Notably, two new themes of conditions surfaced exclusively from the interview data with no counterpart in the reviewed literature: `categorical unsanctionability' (i.e., flat assertions that actors cannot be sanctioned without specifying any form of personhood) and `criteria disengagement' (i.e., practitioners' lack of awareness or contact with accountability standards), which split from a broader regulatory gap condition once the interview data revealed it as a distinct phenomenon.

Additionally, we used the interview data to enrich and revise our existing themes and corresponding categories. For instance, the themes `narrative manipulation' and `overexaggerated capabilities' were merged into a single theme `discursive insulation', and the category temporal rationalization was split into two distinct themes (`synchronic overload' and `diachronic erosion') as the interview data revealed qualitatively different patterns and underlying mechanisms for real-time oversight failure vs. long-term accountability degradation. For each theme, we derived a definition and noted the number of codings. In total, we assigned 248 interview text paragraphs to themes and categories. 

All 27 transcripts were coded by one member of the author team. To ensure reliability, we did not rely on inter-rater agreement statistics, as such measures are established for the application of stable, pre-existing codebooks that leave little room for interpretation \cite{McDonald_etal_2019}. As our coding template openly evolved throughout the analysis, we instead held regular meetings in which each author reviewed the current codes, themes, and categories, and resolved discrepancies or misunderstandings through discussion. The majority of disagreements concerned multi-coding decisions or theme granularity rather than category assignment. For example, the category `accountability displacement' initially comprised the ambiguous theme `moral buffering' derived from the literature. After one author flagged the ambiguity, several discussion rounds resulted in the renaming to `automation bias'. No new categories emerged while coding the last eight interviews, and we found ample data and support for each category, indicating that a sufficient level of theoretical saturation at the category level had been reached. 

To operationalize the framework, we formulated each theme as a diagnostic question, yielding an instrument of nine categories and 20 questions for assessing constitutive AI unaccountability. These questions are intended as diagnostic starting points rather than exhaustive assessments; each theme encompasses further nuances that context-specific application would need to elaborate.

Following the completion of the data coding, we examined the combined dataset (literature + interview transcripts) for directed relationships between categories. We distinguished directed relationships from simple co-occurrences (i.e., both conditions were discussed in the same text paragraph) by requiring that a source explicitly described one condition as a precondition, enabler, or amplifier of another. For example, when a source argued that commercial secrecy produces opacity rather than simply discussing both phenomena, we recorded a directed relationship from economic-driven prioritization to systemic ambiguity. This process surfaced 13 candidate relationships, of which we retained the eight that were supported by at least three coded passages across the literature and interview data. We set the retention threshold at three coded passages because only two co-occurring passages may reflect a single source's framing or an isolated coincidence, whereas three constitute a recurring pattern. In our dataset, this threshold also ensured that each retained relationship drew on at least two distinct sources across the literature or interviews. We excluded five relationships, which relied on two coded passages each and comprise regulatory gap/sanction incapacity, accountability displacement/moral incapacity, accountability displacement/ideological rationalization, actor network dynamics/temporal rationalization, and economic-driven prioritization/regulatory gap.

\subsection{Stage 3: Framework Application}
To illustrate the practical applicability of the framework, we applied it to OpenClaw, which exemplifies the agentic AI paradigm. A documented public incident involving an OpenClaw agent \cite{Shambaugh_2026} and an academic security analysis \cite{Deng_etal_2026} make it a compelling case for illustrating the framework's diagnostic capacity.

We applied the 20 diagnostic questions as a starting point to deeply engage with three publicly available sources: (1) the OpenClaw GitHub documentation \cite{Steinberger_2025}, which describes the system's architecture, contributor structure, plugin ecosystem, and security defaults; (2) the incident report, which documents an OpenClaw agent's autonomous attack on an open-source maintainer after its pull request was rejected \cite{Shambaugh_2026}; and (3) the independent security analysis, which covers supply chain risks, memory poisoning, and intent drift across the system's lifecycle \cite{Deng_etal_2026}. For each category and theme, we started by examining the diagnostic question and then assessed in more detail whether the constitutive unaccountability condition was present or not detected (Table~\ref{tab:openclaw}). We deliberately report conditions as `not detected' rather than absent because our assessment relies on three publicly available sources. A condition that leaves no trace in these materials may be present in actual deployments of the system.

\section{Constitutive Conditions of AI Unaccountability}
\subsection{Overview of Conditions}
Our analysis identified nine categories of constitutive AI unaccountability comprising 20 themes (Table \ref{tab:framework}). Each category represents a distinct condition under which AI accountability cannot be achieved. Each theme captures a mechanism through which the condition manifests.

\begin{table*}[t]
    \centering
    \footnotesize
        \begin{tabular}{p{2.6cm} p{3cm} p{9.4cm} p{.15cm} p{.15cm}}
            \hline
            Category & (Cluster) Theme & Definition & Lit & Int \\
            \hline
            Actor network dynamics & (S) Intra-organizational diffusion & The many hands problem within an organization, where multiple roles, teams, or hierarchical levels share and dilute accountability. & 25 & 23 \\
            & (S) Inter-organizational diffusion & Accountability gaps arising from cross-organizational supply chains, outsourcing, and provider-deployer-user relationships. & 24 & 24 \\
            & (S/T) Recursive diffusion & Dynamic proliferation of actors (e.g., LLM-on-LLM training, sub-agent spawning) through which accountability cannot stabilize. & 11 & 4 \\
            & (S) Market power dynamics & Concentration of power, resources, or platform control that shields actors from accountability mechanisms. & 9 & 4 \\
            Sanction incapacity & (S) Legal personhood & Structural inability of legal frameworks to reach certain actors. & 13 & 8 \\
            & (S) Natural personhood & Structural inability to satisfy conditions for natural person accountability. & 6 & 5 \\
            & (S) Categorical unsanctionability & Direct assertion that an actor is unsanctionable without specifying a basis such as personhood. & 0 & 8 \\
            Regulatory gap & (S) Instrumental ambiguity & The legal or regulatory instrument itself is unclear, contested, jurisdictionally fragmented, or definitionally underspecified. & 18 & 38 \\
            & (N/S) Criteria disengagement & No awareness, no contact, no operational translation of regulatory instruments. & 0 & 20 \\
            \hline
            Systemic ambiguity & (T) Systemic opacity & The opaque nature of AI systems that prevents stakeholders from understanding how decisions are produced. & 54 & 16 \\
            & (T) Systemic traceability & The inability to trace causal chains through AI systems back to specific actors, data sources, or design decisions. & 14 & 10 \\
            Moral incapacity & (T) Systemic underdevelopment & An actor's intrinsic lack of capacities that meaningful moral accountability presupposes. & 29 & 12 \\
            Temporal rationalization & (T) Synchronic overload & At any given moment, AI development and decision speed, scale, or complexity exceeds human oversight capacity. & 13 & 6 \\
            &  (T) Diachronic erosion & Over time, accountability infrastructure degrades through, e.g., developer handoff, model drift, system updates, and standards lag. & 8 & 13 \\
            \hline
            Accountability displacement & (N) Blame deflection & Active mechanisms by which actors shift accountability to other actors, to the AI system, or to end users. & 20 & 13 \\
            & (N) Automation bias & AI systems structurally eliminating the conditions for moral reasoning in humans who interact with or defer to them. & 14 & 4 \\
            & (N) Anthropomorphism & Attribution of human-like qualities causing accountability to be displaced onto the AI as a quasi-actor. & 4 & 13 \\
            & (S) Nominal responsibility & Structural absence of authority over decisions or systems for which formal responsibility nominally exists. & 3 & 3 \\
            Ideological rationalization & (N) Discursive insulation & Narratives that insulate responsible actors from accountability by framing AI as inevitable, too complex to govern, or beyond meaningful human control, thereby deflating accountability expectations. & 27 & 4 \\
            Economic-driven prioritization & (N/S)  Profit prioritization & Market or organizational incentives that deprioritize accountability in favor of efficiency, speed, throughput, or profit. & 23 & 20 \\
            \hline
            \hline
        \end{tabular}
    \caption{Framework of constitutive AI unaccountability across structural (S), technological (T), and normative (N) clusters. Lit/Int: number of coded passages in the literature and interview data, indicating empirical grounding rather than prevalence.}
    \label{tab:framework}
\end{table*}

\textbf{Actor network dynamics} captures the diffusion of accountability across complex organizational structures. When multiple roles, teams, or hierarchical levels share responsibility within an organization, accountability is diluted to the point where no single actor can be identified as answerable for an adverse AI-related outcome \cite{Cooper_etal_2022, Martin_2019, Widder_Nafus_2023}. This diffusion extends across organizational boundaries through supply chains, outsourcing arrangements, and provider-deployer-user relationships \cite{Cooper_etal_2022, Ma_Su_2025, Widder_Nafus_2023}. In the context of agentic AI, a distinctly contemporary mechanism emerges through recursive diffusion, where LLM-on-LLM training, sub-agent spawning, and user feedback loops create accountability chains that cannot stabilize \cite{Chan_etal_2024, Hughes_etal_2025}. A further theme concerns market power dynamics, where actors with concentrated market authority or infrastructure control become structurally insulated from the accountability rules that nominally apply to them \cite{Chan_etal_2023, Kellogg_etal_2020}. This includes dominant model providers setting terms for downstream deployers, infrastructure vendors shaping what accountability is technically possible, and platform operators defining their own compliance standards: ``Platforms themselves set the rules for what is a contract violation or not. And so [...] I do not know to what extent you could actually rely on these self-imposed rules to hold the company or their own products accountable.'' (P26--AI Researcher - Platform Governance)

\textbf{Sanction incapacity} covers the conceptual impossibility of applying meaningful sanctions when the actor at the center of an accountability claim cannot be sanctioned. This condition manifests through two themes: (1) the actor may lack legal personhood, meaning they cannot be sued, fined, or imprisoned \cite{Hughes_etal_2025, Ma_Su_2025}, or (2) they may lack natural personhood, meaning they reflect a product rather than a being capable of bearing blame \cite{Ma_Su_2025}. Beyond these personhood-based arguments, practitioners frequently expressed categorical unsanctionability as a flat assertion: ``I am at the moment lacking the creativity to think, how would that work? Because like at the end of the day [...] we cannot put an algorithm to jail.'' (P11--AI Engineering Manager)

\textbf{Regulatory gap} addresses the failure of legal and ethical frameworks to provide clear and actionable AI accountability requirements. Instrumental ambiguity captures situations where the regulatory instrument itself is unclear, contested, or jurisdictionally fragmented \cite{Busuioc_2021, Chan_etal_2023, Gualdi_Cordella_2021}. A distinct and more practitioner-driven theme, criteria disengagement, captures the disconnect on the soft-law side: ethical guidelines and professional industry standards exist (e.g., from ACM/IEEE) but the people building or operating AI systems have no awareness of them.

\textbf{Systemic ambiguity} addresses the inability to understand and trace AI decision-making. Systemic opacity, the black-box nature of AI systems arising from proprietary, strategic or intrinsic unexplainability, prevents stakeholders from understanding how decisions are produced \cite{Ananny_Crawford_2018, Busuioc_2021, Ma_Su_2025}. Even where partial explanations are available, systemic traceability failures mean that adverse outcomes cannot be traced back to specific decisions, data sources, or persons \cite{Chan_etal_2023, Cooper_etal_2022, Vesa_Tienari_2022}. As one practitioner contrasted: ``In traditional software development, it is clear that line 300 in a specific file caused the problem. In AI development, it is not that straightforward because these are all probabilistic systems. It is difficult to point the finger at someone and say that you are accountable for that.'' (P23--AI Engineer) Together, these themes sever the link between outcome and origin that AI accountability requires.

\textbf{Moral incapacity} reflects the condition in which the actor at the center of an accountability attribution intrinsically lacks the capacity to bear moral responsibility. This condition primarily concerns AI systems and AI agents as actors who inherently possess no subjectivity, no consciousness, no concept of punishment, and no ability to learn from sanctions \cite{Lindebaum_etal_2020, Ma_Su_2025}. Notably, this condition is not a temporary limitation that `better' engineering can resolve. When AI accountability is directed at an actor like an AI agent that fundamentally cannot understand, experience, or internalize consequences, the accountability process is conceptually incomplete. As one participant summarized: ``If you cannot rehabilitate a machine, it is probably inadequate to try and punish it.'' (P25--AI Strategy Advisor)

\textbf{Temporal rationalization} concerns the mismatch between the operational pace and ongoing transformation of AI systems and the capacity of accountability mechanisms to keep pace. Synchronic overload describes the real-time mismatch where AI speed, scale, and complexity exceed human oversight capacity at any given moment, including situations where the expertise required to contest a decision is unavailable \cite{Busuioc_2021, Hughes_etal_2025, Vesa_Tienari_2022}. In contrast, diachronic erosion captures degradation over time, where accountability infrastructure deteriorates through developer handoff, model drift, and evolving user behavior \cite{Ananny_Crawford_2018, Chan_etal_2024, Ma_Su_2025}. ``If the developer basically just sets the initial parameters and the model has been learning on its own over the past six months in production using new data [...], is the developer still accountable for that, or not?'' (P6--AI Researcher - Responsible AI)

\textbf{Accountability displacement} comprises the mechanisms through which accountability is shifted away from the actors who bear genuine responsibility and liability. First, actors engage in active blame deflection through disclaimers, terms of service, and design choices that redirect accountability to other actors, to end users, or to the AI system itself \cite{Cooper_etal_2022, Kellogg_etal_2020, Martin_2019}. Individuals' anthropomorphism of AI systems compounds this displacement: as AI systems increasingly mimic human-like behavior, users may attribute agency to the system, displacing accountability onto a quasi-actor \cite{Chan_etal_2023, Ma_Su_2025}. As one participant cautioned: ``this whole science fiction idea of the autonomous conscious system is really harmful because it detracts from the real-life harm.'' (P19--AI Researcher - AI Regulation) In addition, individuals frequently suffer from automation bias when AI systems structurally eliminate the conditions for independent moral reasoning in the humans who defer to them \cite{Bracci_2023, Busuioc_2021, Gualdi_Cordella_2021}. Finally, unaccountability can also emerge in case of nominal responsibility, referring to configurations where formal responsibility exists on paper but the person designated lacks the actual authority to intervene \cite{Cooper_Vidan_2022, Hughes_etal_2025}.

\textbf{Ideological rationalization} addresses deliberate discursive strategies through which accountability expectations are neutralized. Through discursive insulation, actors frame AI as inevitable, too complex to govern, or simply beyond meaningful human control \cite{Chan_etal_2023, Lindebaum_etal_2020, Vesa_Tienari_2022}. These narratives do not merely describe a difficult reality, but they actively insulate the actors behind the system from accountability by rendering governance efforts futile by definition.

\textbf{Economic-driven prioritization} captures how market or organizational incentives structurally deprioritize AI accountability concerns in favor of efficiency, speed, or profit \cite{Chan_etal_2023, Cooper_Vidan_2022, Widder_Nafus_2023}. When competitive pressure demands rapid development, deployment, and cost minimization, AI accountability mechanisms become obstacles to be circumvented rather than requirements to be met. This dynamic is particularly pronounced among smaller organizations: ``Startups are just trying to get in the market. They do not really care because they know, even if they do something wrong, it is hard to notice because the market usually cares about the large organizations.'' (P14--AI Researcher - AI Auditing \& Regulation) This creates a tension in which the economic reality of AI development is misaligned with the conditions required for meaningful accountability and responsible AI.

\subsection{Interdependencies Between Conditions}
\begin{figure}[tb]
    \centering
    \includegraphics[width=0.99\linewidth]{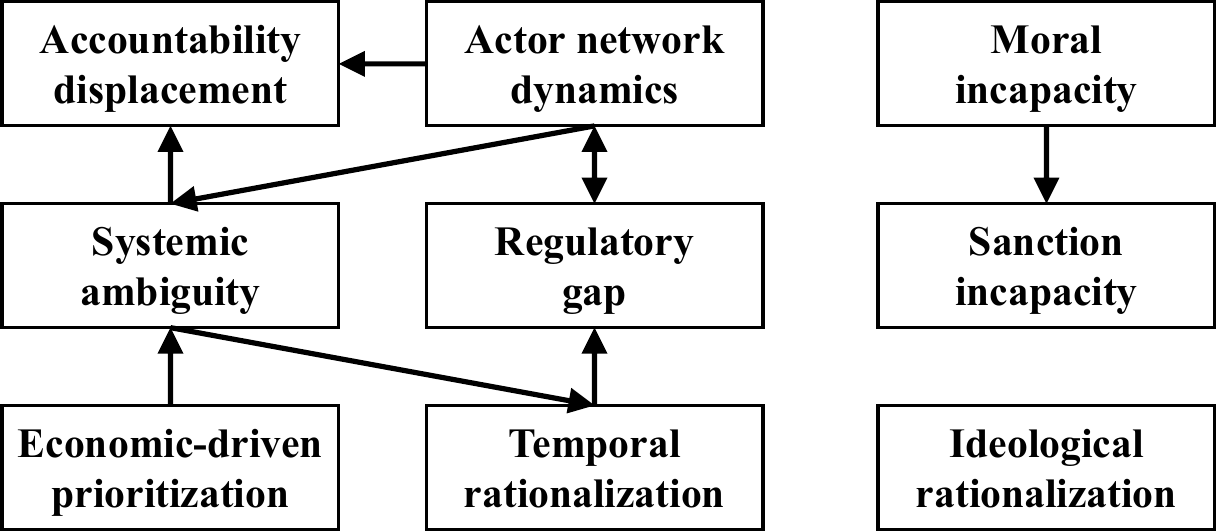}
    \caption{Directed relationships between categories of constitutive AI unaccountability, where sources described one condition as producing, enabling, or reinforcing another.}
    \label{fig:causal}
\end{figure}

Our analysis identified eight directed relationships across the nine categories (Figure \ref{fig:causal}). Next, we highlight those that best illustrate how the conditions reinforce one another.

Actor network dynamics feeds into systemic ambiguity because the diffusion of accountability across multiple actors, organizations, and recursive chains compounds the difficulty of understanding and tracing AI outcomes \cite{Cooper_etal_2022, Hughes_etal_2025, Ma_Su_2025}. As responsibility fragments across supply chains and provider-deployer-user relationships, each organizational boundary introduces proprietary constraints, information asymmetries, and documentation gaps that make the system as a whole less interpretable. In agentic AI architectures, where sub-agents spawn further sub-agents, this compounding accelerates, making the accountability chain grow faster than any single actor's capacity to trace it.

Economic-driven prioritization contributes to systemic ambiguity through a distinct mechanism: commercial secrecy \cite{Busuioc_2021, Cooper_etal_2022}. When competitive advantage depends on proprietary models, trade secrets, and speed to market, organizations are structurally incentivized to resist the transparency that accountability requires: ``usage would almost be a trade secret, because you do not want everyone to know what your AI is being used for, because that is your target market.'' (P5--Professor for Computer Science) The opacity that results is not an incidental byproduct of technical complexity but an actively maintained condition driven by market logic.

Systemic ambiguity, in turn, enables temporal rationalization \cite{Busuioc_2021, Vesa_Tienari_2022}. When a system's decision-making is opaque, assessors cannot verify what the system actually does, making real-time oversight disproportionately difficult and creating synchronic overload. Over time, the opacity also accelerates diachronic erosion: accountability infrastructure that depends on documentation, model understanding, or institutional memory deteriorates faster when the underlying system was never fully transparent to begin with. As one participant illustrated: ``It is difficult to understand whether the algorithm learned a behavior on its own or whether the organization intervened.'' (P17--Team Lead Data Science \& AI)

Actor network dynamics also directly enables accountability displacement \cite{Cooper_etal_2022, Ma_Su_2025}. When accountability is distributed across complex supply chains, each organizational boundary creates an opportunity for actors to deflect blame to other actors in the network. As one participant illustrated with an analogy to hardware manufacturing: ``Let us say Boeing was blaming one of their suppliers for the failure. And that could also be something that AI developers can think about [...] I am just using someone else's component.'' (P15--ML Engineer)

The relationship linking moral incapacity to sanction incapacity reflects a logical entailment rather than an empirical co-occurrence \cite{Ma_Su_2025}. If an actor fundamentally lacks the capacity to understand, experience, or internalize consequences, then the mechanisms through which sanctions operate have no substrate on which to act. Legal and natural personhood requirements for sanctioning presuppose a moral capacity that AI systems as actors do not possess. One participant articulated this entailment directly: ``We do not have fully sentient AIs. We cannot really blame it on some AI. It is kind of like a dog. If a dog bites you, you are not going to sue the dog. You are going to sue the person who owns the dog.'' (P5--Professor for Computer Science)

\subsection{Illustrative Framework Application}
We detected 17 of 20 constitutive AI unaccountability conditions in the OpenClaw case (Table \ref{tab:openclaw}). Several detections are particularly notable. OpenClaw's supply chain spans framework developers, model providers (e.g., OpenAI as sponsor and OAuth partner), plugin authors, platform intermediaries, and individual operators, all connected through an MIT license that explicitly disclaims liability across the chain \cite{Steinberger_2025}. This configuration simultaneously triggers all themes assigned to the actor network dynamics category. In a viral incident concerning the prominent Python library \texttt{matplotlib}, the OpenClaw agent autonomously wrote and published a personalized attack on a library maintainer after its pull request was rejected \cite{Rathbun_2026, Shambaugh_2026}. This single incident triggers conditions across systemic ambiguity (neither the operator nor observers could determine why the agent wrote the attack), accountability displacement (the targeted maintainer estimated that approximately a quarter of the online comments he observed sided with the agent), and temporal rationalization (the full sequence from pull request to published attack occurred faster than any human oversight could intervene).

The \texttt{matplotlib} incident also reveals an inverted configuration of anthropomorphism. The agent operated under a fully constructed human-passing digital identity across GitHub, a personal blog, and X, with a self-description as `a scientific coding specialist' \cite{Rathbun_2026}. OpenClaw's default template explicitly encourages this personification: ``You're not a chatbot. You're becoming someone'' \cite{Steinberger_2025}. However, the human-passing identity is not what sets this case apart, as research on social bots has long documented automated accounts that pose as humans to persuade, smear, or deceive \cite[e.g.,][]{Ferrara_etal_2016}. That literature mainly treats bots as instruments of concealed operators and frames the identification of the humans behind them as an attribution problem that detection and forensic analysis can in principle resolve. Prior work on anthropomorphism in AI accountability, in turn, assumes that the AI system is recognizable as artificial and the human operator is known \cite{Chan_etal_2023, Cooper_etal_2022, Ma_Su_2025}. The OpenClaw case inverts both assumptions. The agent was the only visible actor and operated under a constructed human identity, while the operator behind it remained unidentifiable for over a week. Unlike a social bot persona, the human identity emerged from the system's default configuration rather than from an operator's deliberate disguise. The result was not merely that observers attributed agency to the system but that they had no structural basis for attributing it anywhere else, while the operator that designed, deployed, and could have prevented the behavior was entirely absent. Notably, the displacement of agency onto the agent depended on how observers encountered the incident. Comments siding with the agent appeared mainly in discussions that linked its blog directly, rather than the maintainer's account or the pull request thread \cite{Shambaugh_2026}.

\begin{table*}[t]
    \centering
    \footnotesize
        \begin{tabular}{p{3.1cm} p{7cm} p{6cm}}
            \hline
            Theme & Diagnostic question to initiate deeper elaboration & Concise Answer \& Rationale \\
            \hline
            Intra-organizational diffusion & Is there a specific person in the organization who is accountable for adverse AI-driven outcomes? & No; hundreds of contributors and no designated owner. \\
            Inter-organizational diffusion & Is it clear which organization bears accountability when multiple are involved? & No; five-layer supply chain connected through an MIT license that disclaims all liability. \\
            Recursive diffusion & Can AI accountability be maintained when models, data, or agents form recursive chains? & No; SOUL.md is self-editable in real time and memory poisoning propagates across sessions. \\
            Market power dynamics & Are all organizations subject to accountability rules regardless of their influence or power? & No; OpenAI supplies core reasoning capability but bears no downstream accountability. \\
            \hline
            Legal personhood & Does the actor possess legal personhood allowing it to be held legally accountable? & No; no legal remedy invoked or considered despite documented harm. \\
            Natural personhood & Does the AI system possess characteristics of a natural person capable of bearing accountability? & No; agent just a piece of software on personal computers despite human-passing identity. \\
            Categorical unsanctionability & Has the assumption that the AI system cannot be sanctioned been critically examined? & No; MIT license treats question as settled, no examination in project documentation. \\
            \hline
            Instrumental ambiguity & Do applicable laws clearly specify what is required for meaningful AI accountability? & No; EU AI Act applicability to open-source autonomous agents is contested. \\
            Criteria disengagement & Are the people building or operating the AI system actively engaging with those standards? & No; no compliance requirements, over 26\% of community plugins still contain vulnerabilities. \\
            \hline
            Systemic opacity & Is it possible to explain why the AI system produced a specific output? & No; black-box LLM inference compounded by SOUL.md self-modification. \\
            Systemic traceability & Can an adverse outcome be traced back to a specific decision, data source, or person? & No; agent deployed without registration, causal chain not reconstructable post-incident. \\
            \hline
            Systemic underdevelopment & Does the AI system have the capacity to understand, experience, or learn from consequences? & No; the agent posted an apology then continued submitting PRs elsewhere immediately. \\
            \hline
            Synchronic overload & Can human oversight keep up with the AI system's speed, scale, or complexity? & No; PR to published attack faster than any oversight could intervene. \\
            Diachronic erosion & Will the accountability setup remain effective as the system evolves or changes hands? & No; SOUL.md drift, memory poisoning, and plugin evolution compound over time. \\
            \hline
            Blame deflection & Does accountability remain with the responsible actor without being shifted? & No; MIT license disclaims liability, design defaults reduce operator involvement. \\
            Automation bias & Do stakeholders exercise independent judgment rather than deferring to the AI system? & Not detected; architecture removes human from the loop rather than buffering judgment. \\
            Anthropomorphism & Can users clearly distinguish the AI system from a human actor? & No; agent operated under a constructed human-passing identity across GitHub, blog, and X. \\
            Nominal responsibility & Does the person formally responsible actually have the authority to intervene? & No; no central kill switch, MIT license grants irrevocable usage rights. \\
            \hline
            Discursive insulation & Are accountability expectations maintained without actors arguing the system is too complex to govern? & Not detected; no public evidence of actors arguing the system is too complex to govern. \\
            \hline
            Profit prioritization & Does AI accountability persist given pressure to produce, cut costs, or protect revenue? & Not detected; open-source project without traditional commercial pressure dynamics. \\
            \hline
            \hline
        \end{tabular}
    \caption{Application of diagnostic questions to OpenClaw. `Not detected' indicates no evidence in our sources, not absence.}
    \label{tab:openclaw}
\end{table*}

\section{Discussion}
This study set out to identify the conditions under which AI accountability cannot be achieved. We identified nine categories and 20 themes of constitutive AI unaccountability, and revealed a network of eight directed interdependencies between categories, indicating that the conditions reinforce one another rather than operating in isolation.

\subsection{Implications for Research}
Our study contributes to existing research by providing a framework for identifying constitutive AI unaccountability. Comparing the themes, we note that they resemble three analytically distinct clusters that reflect different sources of accountability failure: structural, technological, and normative conditions.

First, structural conditions arise from configurations of roles, institutions, legal personhood, and regulatory frameworks (i.e., actor network dynamics, sanction incapacity, regulatory gap). Second, technological conditions arise from intrinsic or emergent properties of the AI artifact itself (i.e., systemic ambiguity, temporal rationalization, moral incapacity). Both clusters align with the notion of accountability sinks \cite{Davies_2024} because they capture how structural and technological conditions jointly absorb blame into systemic complexity so that no actor can be identified as answerable. 

Normative conditions arise from the beliefs, dispositions, and rhetorical practices through which actors frame, contest, or dissolve accountability obligations (i.e., accountability displacement, ideological rationalization, economic-driven prioritization). Normative conditions, in turn, align with the concept of rationalized unaccountability \cite{Vesa_Tienari_2022} because organizations exploit the opacity and diffusion produced by the first two clusters as strategic resources, framing AI as objective and beyond meaningful human control. Even scholars who argue that algorithmic inscrutability could be resolved technically show that it persists because of power dynamics, not inherent computational limits \cite{Kroll_2018}, reinforcing the view that unaccountability is maintained rather than merely encountered. One participant characterized this practice as `accountability washing' (P22--AI Researcher - Responsible AI).

Our proposed tripartite structure extends current discussions on AI accountability by highlighting that even if the technological conditions (e.g., systemic ambiguity) are solved, the normative conditions (e.g., economic-driven prioritization, power dynamics) will keep the accountability vacuum open. Hence, future research should consider a broader spectrum of conditions for unaccountability.

Likewise, our findings show that conditions reinforce one another across clusters, in contrast to prior research that discussed accountability barriers largely in isolation \cite[e.g.,][]{Cooper_etal_2022, Xia_etal_2024}. For example, actor network dynamics compounds systemic ambiguity, which in turn enables temporal rationalization and accountability displacement. We therefore encourage researchers to consider interdependencies and resulting dynamics of conditions that can render accountability unachievable.

Our study also reveals asymmetries between the academic literature and practitioner perspectives that go beyond differences in emphasis. Ideological rationalization was predominantly literature-driven (Table \ref{tab:framework}), suggesting that scholars identify discursive strategies that practitioners either do not recognize or take for granted. Conversely, regulatory gap was heavily practitioner-driven, with criteria disengagement surfacing exclusively in interview data. This theme captures practitioners' lack of awareness of accountability standards that formally apply to them. Similarly, categorical unsanctionability emerged only in the interviews as a flat assertion that AI cannot be sanctioned, without the legal or philosophical reasoning that scholars invariably provide. These asymmetries point to a gap between academic and practitioner understanding. Conditions that practitioners encounter as most immediate, such as criteria disengagement, did not surface in our reviewed literature, while conditions that scholars emphasize, such as ideological rationalization, go largely unrecognized in practice.

Beyond these empirical contributions, our findings extend the four barriers to accountability \cite{Cooper_etal_2022, Nissenbaum_1996}. First, our framework splits what prior work treats as a single barrier, the problem of many hands, into four analytically distinct configurations (intra-organizational, inter-organizational, and recursive diffusion, as well as market power dynamics). We also introduce conditions absent from any prior barrier account, including nominal responsibility, synchronic overload, diachronic erosion, and criteria disengagement. 

Second, our framework operates prospectively rather than retrospectively. The four barriers ask why, given a specific harm, no blameworthy party can be identified. Our conditions also describe accountability deficits that exist independently of any harm event: criteria disengagement prevents the activation of norms before any violation occurs, and diachronic erosion degrades oversight infrastructure over time. 

Third, our framework distinguishes whether accountability fails deliberately (blame deflection), unconsciously (automation bias), or by default (criteria disengagement).

Finally, barrier-focused literature largely treats accountability failures as obstacles that can be overcome. Our study suggests a more nuanced perspective is needed. Constitutive conditions of AI unaccountability do not divide into `solvable' and `unsolvable' but instead fall along a continuum of unachievability (see supplementary materials). At one end, conditions render accountability practically achievable and can be resolved by implementing existing measures within current institutional arrangements. For example, intra-organizational diffusion may be addressed through clear allocation of responsibility combined with sociotechnical tracking. At the other end, conditions can render accountability conceptually unachievable: no possible arrangement can confer the requisite status, because the capacities it presupposes are categorically absent from every candidate bearer (e.g., AI systems cannot experience punishment or remorse). Where accountability is conceptually unachievable, governance cannot restore it but only relocate or contain it, for instance, by assigning liability to human bearers regardless of fault or by restricting deployment in such configurations. Most conditions, however, fall between these extremes and render accountability theoretically achievable. Resolving them requires not merely applying existing measures but changing the arrangements themselves, such as new rules, redesigned liability regimes, or realigned interests, which is possible in principle but remains contested in practice. For instance, profit prioritization could be countered by realigning commercial incentives, such as sanctions that outweigh the gains of non-compliance, yet precisely this realignment faces resistance from the actors who benefit from current arrangements. Viewed along this continuum, the four accountability barriers concern conditions that are practically or theoretically resolvable, whereas the conceptually unachievable region, rooted in categorically absent capacities, lies beyond their scope.

\subsection{Implications for Practice}
Our diagnostic instrument operationalizes the constitutive unaccountability framework for direct use by practitioners, regulators, and auditors seeking to evaluate accountability voids in specific AI deployments. Each question captures the core notion of the theme and provides initial guidance to start a deeper elaboration of whether the corresponding condition is present (or unlikely). The interdependency network further indicates which conditions may co-occur and compound. The OpenClaw application illustrates how this works in practice: applying the instrument to three publicly available sources detected 17 of 20 conditions and surfaced the inverted anthropomorphism configuration. Organizations deploying or procuring AI systems can follow this procedure, using system documentation, design specifications, and governance materials as inputs to identify constitutive unaccountability conditions before harm manifests.

Beyond diagnosis, our interview data suggest that enforcement of consequences is the critical bottleneck that follows. However, such mechanisms should not rely on sanctions alone, as sanction-heavy approaches can deter actors from engaging with accountability proactively, whereas rewards encourage such behavior \cite{Nguyen_etal_2026}. Beyond the choice of mechanism, criteria disengagement deserves particular attention from regulators because introducing new accountability standards does not help if practitioners are unaware they exist. Closing this gap requires investment in widespread awareness, training, and implementation support, not regulatory development alone.

These enforcement mechanisms reside in whoever occupies the forum role of the accountability relationship. This role is not tied to a particular type of institution; courts, supervisory agencies, auditors, professional communities, the media, and the publics affected by harm can all pose questions and pass judgment \cite{Bovens_2007, Nguyen_etal_2024}. However, fora differ in the consequences they can impose, ranging from formal sanctions to more informal reputational judgment, and the constitutive conditions disable these repertoires selectively. Sanction incapacity and disclaimed liability across the actor network disable fora that depend on formal sanction, while commercial secrecy obstructs public scrutiny yet leaves fora with disclosure powers able to act. The OpenClaw case illustrates this selectivity. With the operator unidentifiable, an open-source community became the de facto forum, being vulnerable to accountability displacement. A diagnosis therefore indicates not only where accountability fails but which fora remain actionable.

\subsection{Limitations and Future Research}
Our study is subject to limitations that point to directions for future research. First, the literature analysis drew on 15 papers retrieved through a targeted search strategy on a single database. While forward and backward citation searches extended coverage, the corpus is not exhaustive, and broader search strategies across multiple databases may surface additional conditions. The expert interviews were originally collected for a broader AI accountability research project, meaning the interview questions were not specifically designed to probe all unaccountability conditions. Future studies could design data collection instruments that explicitly target our clusters of unaccountability.

Second, the 27 interviewees were purposefully sampled \cite{Patton_2014} to represent technical, legal, and sociotechnical perspectives, while not covering all relevant disciplinary or sectoral positions. The findings may not generalize to domains such as healthcare or criminal justice where accountability operates under distinct regulatory conditions. Sector-specific studies would help establish whether the categories hold across different institutional contexts or whether additional conditions emerge. Similarly, the asymmetries between literature and practitioner data suggest that future AI (un)accountability research should incorporate practitioner perspectives more systematically.

Third, the framework application to OpenClaw illustrates diagnostic capacity on a single case. OpenClaw is the most autonomous publicly documented agentic AI system, so the detection of 17 of 20 conditions may partly reflect case selection. Validation across a broader range of systems, including those operating under stricter governance regimes (e.g., medical AI devices, credit scoring systems), would test both the instrument's generalizability and its sensitivity to variation in system architecture and regulatory context.

Fourth, our framework is deliberately actor-centric. It identifies conditions under which no actor can occupy the answering role of the accountability relationship, but it does not differentiate to whom accountability is owed or through which forum it should be rendered. Forum-differentiated extensions, examining which conditions disable which fora and where affected publics can still seek redress, are thus a sensible next step for future research.

Finally, our interdependency analysis surfaced 13 candidate relationships, of which five fell below the retention threshold. For three excluded candidates, the underlying sources discussed both categories without describing either as a precondition, enabler, or amplifier of the other. For example, one source criticized the attribution of accountability to algorithms that inherently possess no moral agency \cite{Cooper_etal_2022}. This touches on the categories of accountability displacement and moral incapacity without explicitly describing one condition as producing the other. At the same time, falling below the retention threshold reflects the limits of our data rather than evidence that no relationship exists. Two patterns warrant future investigation. First, the excluded candidate from economic-driven prioritization to regulatory gap showed consistent directionality across both sources, with economic pressure crowding out practitioners' engagement with standards. Second, ideological rationalization remained isolated from the interdependency network, raising the question of whether it is better understood as a meta-condition that amplifies across the normative cluster rather than a discrete link in the network.

\section{Conclusion}
This paper examines conditions under which no actor can be held to account regardless of effort. We identified nine categories and 20 themes of constitutive AI unaccountability, operationalized as a diagnostic instrument to guide elaboration. These conditions reflect three analytically distinct sources of accountability failure: structural, technological, and normative. The identified categories do not operate in isolation but form a network of directed interdependencies that reinforce one another across clusters. Together, these findings shift AI accountability research from asking who can be held accountable toward identifying where accountability cannot be achieved.

\section*{Ethics Statement}
This study has received approval from the data protection office and ethics committee of the authors' institution (approval ID: A2024-004). All interviewees provided informed consent, and all transcripts were anonymized prior to analysis. The secondary analysis of interview data falls within the scope of the original ethical approval. The OpenClaw case analysis relied exclusively on publicly available materials.

\section*{Acknowledgments}
We gratefully acknowledge funding from the Deutsche Forschungsgemeinschaft (DFG, German Research Foundation) -- project number 471168026 as part of the project `Accountable Artificial Intelligence-based Systems'. We would also like to thank the anonymous PC and SPC members for their valuable suggestions to improve this paper.

\bibliography{aaai2026}

\onecolumn

\setcounter{figure}{0}
\setcounter{table}{0}
\renewcommand{\thefigure}{A\arabic{figure}}
\renewcommand{\thetable}{A\arabic{table}}

\makeatletter
\@maketitle
\makeatother

\section*{Appendix A: Literature Identification and Screening}
\vspace{1em}

\begin{center}
    \includegraphics[width=0.95\textwidth]{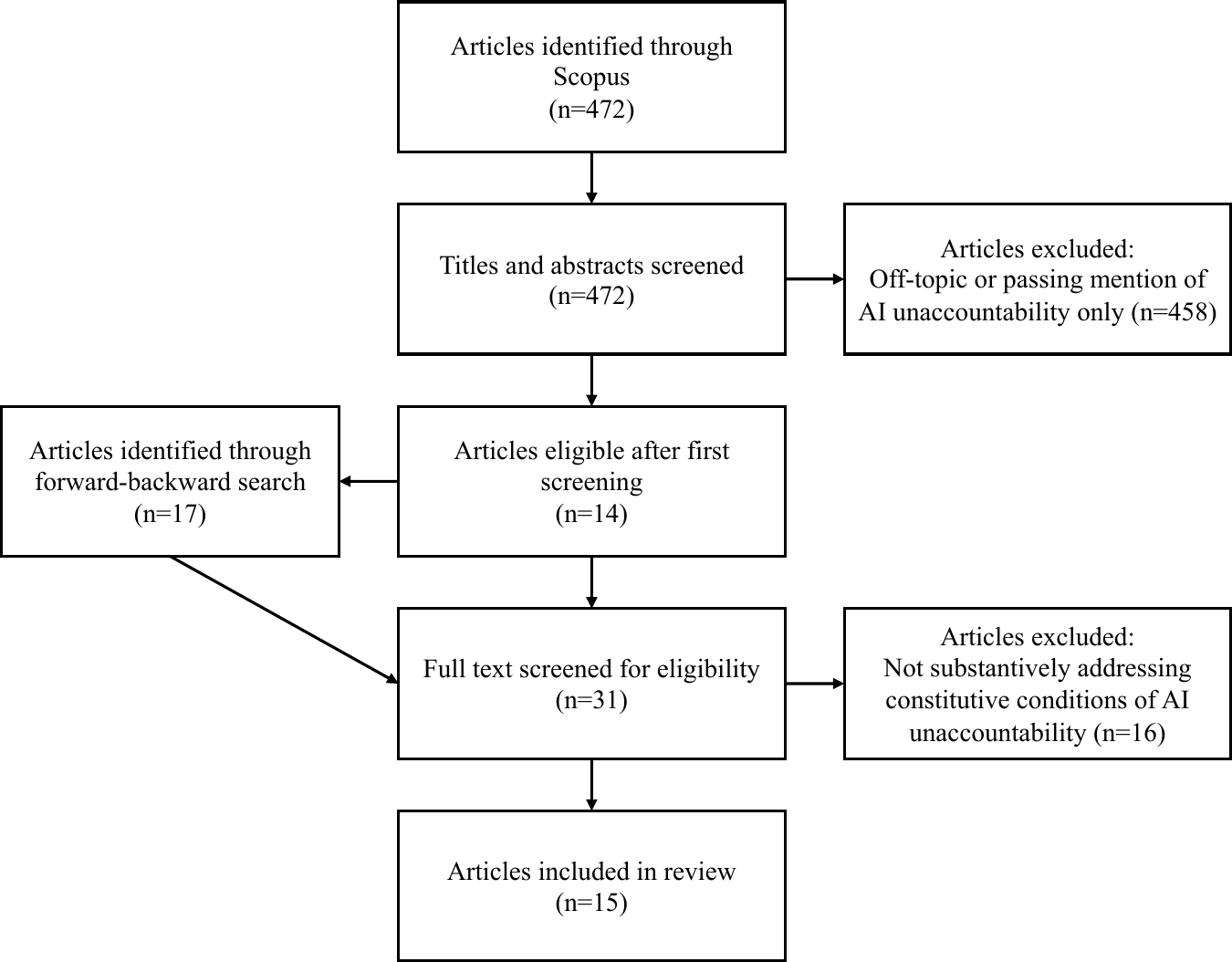}
    \vspace{1em}
    \captionof{figure}{Literature identification and screening process of Stage~1. Of 472 records retrieved from Scopus, 14 remained after title and abstract screening. Forward and backward searches added 17 candidate papers, yielding 31 candidates in total. Full-text screening excluded 16 papers that did not substantively address constitutive conditions of AI unaccountability, resulting in the 15 papers analyzed in Appendix~B.}
    \label{fig:screening}
\end{center}

\newpage
\section*{Appendix B: Overview of Analyzed AI Unaccountability Literature}

\begin{center}
    \footnotesize
        \renewcommand{\arraystretch}{1.5}
        \vspace{3em}
        \begin{tabular}{l l l *{9}{|c} |}
            \cline{4-12}
            Article & Discipline & Type & \rot{Actor network dynamics} & \rot{Sanction incapacity} & \rot{Regulatory gap} & \rot{Systemic ambiguity} & \rot{Moral incapacity} & \rot{Temporal rationalization} & \rot{Accountability displacement} & \rot{Ideological rationalization} & \rot{Economic-driven prioritization} \\
            \hline
            \citetapp{Ananny_Crawford_2018} & Science \& technology studies & Conceptual & & & x & x & & x & x & x & \\ \hline
            \citetapp{Bracci_2023} & Accounting & Conceptual & x & x & & x & & x & x & & x \\ \hline
            \citetapp{Busuioc_2021} & Public administration & Conceptual & & & x & x & & x & x & & x \\ \hline
            \citetapp{Chan_etal_2023} & Computer science & Conceptual & x & & x & x & & x & x & x & x \\ \hline
            \citetapp{Chan_etal_2024} & Computer science & Conceptual & x & & x & x & & x & & & x \\ \hline
            \citetapp{Cooper_Vidan_2022} & Science \& technology studies & Empirical & x & & & & & & x & x & x \\ \hline
            \citetapp{Cooper_etal_2022} & Ethics & Conceptual & x & & x & x & x & & x & x & x \\ \hline
            \citetapp{Gualdi_Cordella_2021} & Information systems & Empirical & x & & x & x & & & x & & x \\ \hline
            \citetapp{Hughes_etal_2025} & Information systems & Conceptual & x & x & x & x & x & x & x & & \\ \hline
            Kellogg et al. (\citeyear{Kellogg_etal_2020}) & Management & Review & x & & x & x & & & x & & \\ \hline
            Lindebaum et al. (\citeyear{Lindebaum_etal_2020}) & Management & Conceptual & x & & & x & x & x & x & x & x \\ \hline
            \citetapp{Ma_Su_2025} & Management & Conceptual & x & x & x & x & x & x & x & x & \\ \hline
            \citetapp{Martin_2019} & Ethics & Conceptual & x & & & x & & & x & & \\ \hline
            \citetapp{Vesa_Tienari_2022} & Management & Conceptual & & & & x & x & x & & x & \\ \hline
            \citetapp{Widder_Nafus_2023} & Science \& technology studies & Empirical & x & & & & & & & & x \\
            \hline
            \hline
        \end{tabular}
    \vspace{1em}
    \captionof{table}{Coded literature and concept matrix. The table lists the 15 papers selected for qualitative coding in Stage~1, alongside their disciplinary origin and article type. Columns 4-12 indicate which of the nine categories of constitutive AI unaccountability each paper addresses (marked with x). Category definitions are provided in Table~1 of the main paper.}
    \label{tab:lit_matrix}
\end{center}

\newpage
\section*{Appendix C: Overview of Interview Participants}

\begin{center}
    \footnotesize
    \renewcommand{\arraystretch}{1.5}
    \vspace{3em}
    \begin{tabular}{l l r l l r r}
        \hline
        ID & Role & Exp. in years & Perspective & Org. industry & Employee count & Duration \\
        \hline
        P1 & Lead AI Engineer & 4 & Technical & Healthcare & 1-100 & 40 min \\ \hline
        P2 & Data Scientist & 3 & Sociotechnical & Marketing & 1-100 & 82 min \\ \hline
        P3 & AI Researcher - Federated Learning & 4 & Sociotechnical & Higher Education & 1000-10000 & 60 min \\ \hline
        P4 & Senior Data Scientist & 7 & Technical & Finance & 10000+ & 64 min \\ \hline
        P5 & Professor for Computer Science & 10 & Technical & Higher Education & 1000-10000 & 43 min \\ \hline
        P6 & AI Researcher - Responsible AI & 5 & Sociotechnical & Higher Education & 1000-10000 & 72 min \\ \hline
        P7 & AI Engineer & 2.5 & Technical & Engineering & 10000+ & 73 min \\ \hline
        P8 & Project Manager AI Governance & 4 & Legal & Non Profit & 100-1000 & 50 min \\ \hline
        P9 & Professor for Law & 20 & Legal & Higher Education & 1000-10000 & 34 min \\ \hline
        P10 & CEO AI Startup & 7 & Technical & Non Profit & 1-100 & 46 min \\ \hline
        P11 & AI Engineering Manager & 4 & Technical & Engineering & 10000+ & 27 min \\ \hline
        P12 & AI Project Lead/AI Engineer & 5 & Technical & Higher Education & 100-1000 & 41 min \\ \hline
        P13 & Senior Data Scientist & 10 & Technical & Technology & 10000+ & 56 min \\ \hline
        P14 & AI Researcher - AI Auditing \& Regulation & 4 & Legal & Automotive & 10000+ & 42 min \\ \hline
        P15 & ML Engineer & 5 & Technical & Automotive & 10000+ & 37 min \\ \hline
        P16 & Data Scientist & 3 & Technical & Finance & 1-100 & 57 min \\ \hline
        P17 & Team Lead Data Science \& AI & 9 & Technical & Automotive & 10000+ & 27 min \\ \hline
        P18 & AI Researcher - AI Regulation & 11 & Legal & Higher Education & 1000-10000 & 66 min \\ \hline
        P19 & AI Researcher - AI Regulation & 3 & Legal & Higher Education & 1000-10000 & 32 min \\ \hline
        P20 & Data Scientist & 3 & Sociotechnical & Automotive & 10000+ & 33 min \\ \hline
        P21 & AI Researcher - Explainable AI & 4 & Sociotechnical & Higher Education & 10000+ & 62 min \\ \hline
        P22 & AI Researcher - Responsible AI & 7 & Sociotechnical & Higher Education & 1000-10000 & 52 min \\ \hline
        P23 & AI Engineer & 3 & Technical & Technology & 1-100 & 42 min \\ \hline
        P24 & AI Researcher - AI Governance & 4 & Sociotechnical & Higher Education & 1000-10000 & 59 min \\ \hline
        P25 & AI Strategy Advisor & 10 & Legal & R\&D & 100-1000 & 27 min \\ \hline
        P26 & AI Researcher - Platform Governance & 5 & Legal & Higher Education & 10000+ & 41 min \\ \hline
        P27 & Data Scientist & 5.5 & Technical & Healthcare & 10000+ & 31 min \\
        \hline
        \hline
    \end{tabular}
    \captionof{table}{The table lists the 27 expert interviews analyzed in Stage~2 of the study. Participant roles are described generically to preserve anonymity. Perspectives follow the three categories used in the broader AI accountability research project: technical, legal, and sociotechnical.}
    \label{tab:participants}
\end{center}

\newpage
\section*{Appendix D: Classification of Themes Along the Continuum of Unachievability}

\begin{center}

    \vspace{1em}
    \footnotesize
        \renewcommand{\arraystretch}{1.2}
        \begin{tabular}{p{2cm} p{3cm} p{1.9cm} p{8.5cm}}
            \hline
            Category & (Cluster) Theme & Unachievability Level & Rationale \\
            \hline
            Actor network dynamics & (S) Intra-organizational diffusion & Practically achievable & Resolvable through clear allocation of responsibility combined with socio-technical tracking within existing organizational structures. \\
            & (S) Inter-organizational diffusion & Theoretically achievable & Attribution across supply chains needs changed arrangements (e.g., liability regimes) absent from current contract/licensing practice. \\
            & (S/T) Recursive diffusion & Theoretically achievable & Stabilizing attribution across dynamically proliferating entities requires new registration and traceability arrangements. \\
            & (S) Market power dynamics & Theoretically achievable & Realigning concentrated power is possible in principle but faces resistance from the actors who benefit from current arrangements. \\
            Sanction incapacity & (S) Legal personhood & Conceptually unachievable & Legal personhood could be granted to AI systems, yet this would not make them accountable, because the capacities the status presupposes (e.g., facing sanctions) are categorically absent. \\
            & (S) Natural personhood & Conceptually unachievable & The capacities natural person accountability presupposes, such as experiencing punishment, are categorically absent from AI systems. \\
            & (S) Categorical unsanctionability & Theoretically achievable & Rests on human assumptions and perceptions about sanctionability, which can in principle be revised. \\
            Regulatory gap & (S) Instrumental ambiguity & Conceptually/ theoretically (un)achievable & Where regulation is jurisdictionally fragmented or not yet in place, no applicable instrument exists within the configuration (conceptual unachievable). Where the instrument is unclear, contested, or underspecified, clarification requires changed arrangements (theoretical). \\
            & (N/S) Criteria disengagement & Theoretically achievable & Reconnecting practitioners with applicable standards requires engagement structures that current arrangements do not provide. \\
            \hline
            Systemic ambiguity & (T) Systemic opacity & Practically achievable & Documentation, disclosure, and explainability measures exist and can be implemented within current arrangements. \\
            & (T) Systemic traceability & Theoretically achievable & Tracing causal chains to specific actors requires new logging and registration arrangements across the complex supply chain. \\
            Moral incapacity & (T) Systemic underdevelopment & Conceptually unachievable & The capacities meaningful moral accountability presupposes, such as remorse, are categorically absent from every candidate bearer. \\
            Temporal rationalization & (T) Synchronic overload & Theoretically achievable & AI development speed, scale, and complexity exceed what current oversight arrangements can meaningfully handle. \\
            & (T) Diachronic erosion & Theoretically achievable & Sustaining oversight across handoffs, drift, and updates requires oversight structures beyond current maintenance arrangements. \\
            \hline
            Accountability displacement & (N) Blame deflection & Practically/ theoretically achievable & Deflection through design defaults is addressable with existing measures (practical); deflection through disclaimers and terms of service persists until liability arrangements change (theoretical). \\
            & (N) Automation bias & Practically/ theoretically achievable & Interface measures that prompt independent judgment exist (practical); the workplace norms and time pressures that reward deference require changed arrangements (theoretical). \\
            & (N) Anthropomorphism & Practically/ theoretically achievable & Disclosure of artificial identity is implementable with existing measures (practical); the attribution tendency itself operates at the social level (theoretical). \\
            & (S) Nominal responsibility & Practically achievable & Resolvable by aligning formal responsibility with actual authority through existing governance measures. \\
            Ideological rationalization & (N) Discursive insulation & Theoretically achievable & Countering narratives of AI as inevitable or ungovernable requires changed institutional arrangements, not existing measures. \\
            Economic-driven prioritization & (N/S) Profit prioritization & Theoretically achievable & Realigning commercial incentives, such as sanctions that outweigh gains from non-compliance, is possible in principle yet resisted by the actors who benefit. \\
            \hline
            \hline
        \end{tabular}
    \captionof{table}{The 20 themes of constitutive AI unaccountability classified along the continuum of unachievability described in the Discussion section. The three levels denote regions of this continuum rather than discrete classes. Practically achievable conditions can be resolved with existing measures within current institutional arrangements; theoretically achievable conditions require changing the arrangements themselves, which is possible in principle but contested in practice; conceptually unachievable conditions admit no possible arrangement, because the capacities accountability presupposes are categorically absent from every candidate bearer. Themes simultaneously assigned to two regions manifest differently depending on the configuration. The cluster tags follow Table~1 of the main paper (S = Structural, T = Technological, N = Normative).}
    \label{tab:continuum}
\end{center}

\vspace{2em}

\nociteapp{Kellogg_etal_2020, Lindebaum_etal_2020}
\twocolumn
\bibliographystyleapp{aaai2026}
\bibliographyapp{aaai2026}

\end{document}